\RequirePackage[bookmarksnumbered,unicode]{hyperref}
\documentclass[sigconf,10pt]{acmart}
\expandafter\def\csname ver@hyperxmp.sty\endcsname{9999/12/31}
\usepackage[table]{xcolor}
\usepackage[export]{adjustbox}
\usepackage{booktabs}
\usepackage{multirow}
\usepackage{subfigure}
\usepackage{mathrsfs}
\usepackage{placeins}
\usepackage{float}
\renewcommand\footnotetextcopyrightpermission[1]{}
\usepackage{graphicx}
\usepackage{amsmath}

\usepackage{amssymb}
\usepackage{bm}
\usepackage{mathtools}
\usepackage{algorithm}
\usepackage{algpseudocode}
\usepackage{enumitem}
\definecolor{pos1}{HTML}{E8F5E9}   % 0–20%
\definecolor{pos2}{HTML}{A5D6A7}   % 20–50%
\definecolor{pos3}{HTML}{4CAF50}   % 50–80%
\definecolor{pos4}{HTML}{1B5E20}   % 80%+

\definecolor{neg1}{HTML}{FFEBEE}   % 0 to -20%
\definecolor{neg2}{HTML}{EF9A9A}   % -20 to -100%
\definecolor{neg3}{HTML}{E53935}   % -100 to -500%
\definecolor{neg4}{HTML}{7F0000}   % -500%+

\newcommand{\dpos}[2]{%  #1=value text, #2=number for threshold
  \ifnum#2>80
    \cellcolor{pos4}\textcolor{white}{\textbf{#1}}%
  \else\ifnum#2>50
    \cellcolor{pos3}\textcolor{white}{#1}%
  \else\ifnum#2>20
    \cellcolor{pos2}{#1}%
  \else
    \cellcolor{pos1}{#1}%
  \fi\fi\fi
}
\newcommand{\dneg}[2]{%  #1=value text, #2=abs number for threshold
  \ifnum#2>500
    \cellcolor{neg4}\textcolor{white}{\textbf{#1}}%
  \else\ifnum#2>100
    \cellcolor{neg3}\textcolor{white}{#1}%
  \else\ifnum#2>20
    \cellcolor{neg2}{#1}%
  \else
    \cellcolor{neg1}{#1}%
  \fi\fi\fi
}
\newcommand{\dzero}{--}
\begin{document}

% \title{Incorporating Multi-view Urban Images with Urban Structure for Socioeconomic Indicators Prediction}
% \title{Incorporating Urban Images with Urban Structure for the Prediction of Multi-level Socioeconomic Indicators}
\title{NetSpatial: Spatially Conditional Traffic Generation for Cellular Planning and Operations}


\author{Shiyuan Zhang}
\affiliation{%
\institution{The University of Hong Kong,\\ Hong Kong, China}
\country{}}
\email{shiyuanzhang@connect.hku.hk}

\author{Jiale Du}
\affiliation{%
\institution{The University of Hong Kong,\\ Hong Kong, China}
\country{}}
\email{u3650171@connect.hku.hk}

\author{Yuanwei Liu}
\affiliation{%
\institution{The University of Hong Kong,\\ Hong Kong, China}
\country{}}
\email{yuanwei@hku.hk}

\author{Kaibin Huang}
\affiliation{%
\institution{The University of Hong Kong,\\ Hong Kong, China}
\country{}}
\email{huangkb@hku.hk}

\author{Hongyang Du}
\authornote{Corresponding author.}
\affiliation{%
\institution{The University of Hong Kong,\\Hong Kong, China}
\country{}}
\email{duhy@hku.hk}

\begin{abstract}
Base station (BS) deployment and operation are fundamental to network performance, yet they require accurate demand understanding, which remains difficult for operators. 
Cellular traffic in dense urban regions is well measured but highly dynamic, which undermines prediction-based management, whereas the scarcity of traffic measurements in emerging regions limits informed deployment decisions.
Existing approaches therefore either depend on manual planning heuristics or use autoregressive predictors that fail to capture stochastic traffic variation.
We present NetSpatial, a unified system for cellular planning and operation through spatially conditional traffic generation. 
NetSpatial exploits multimodal urban context, including satellite imagery and point of interest (POI) distributions, to learn how physical environment and functional semantics shape BS demand. It uses a multi-level flow-matching architecture that separates periodic structure from residual dynamics, enabling direct generation of long-horizon traffic sequences.
NetSpatial supports two complementary decision scenarios, i.e., what-if analysis for deployment planning, which ranks candidate sites using generated traffic profiles, and what-to-do support for network operation, which uses generated traffic forecasts to guide BS sleep scheduling and load balancing.
Experiments on real-world cellular traffic data show that NetSpatial reduces Jensen–Shannon Divergence (JSD) by 29.44\% over the strongest baseline, generalizes across cities in zero-shot experiments, and enables up to 16.8\% energy savings while maintaining over 80\% quality of experience. 
% In dense urban regions, traffic measurements are available but highly dynamic, which makes accurate prediction-based management unreliable, while in emerging regions traffic measurements are scarce, which limits informed deployment decisions.
% A demo is available\footnote{\url{https://anonymous.4open.science/r/Traffic-Flow-Predictor-Demo}} for spatially informed BS planning and operation.
% An interactive demo has been deployed as an online platform\footnote{\url{https://huggingface.co/spaces/Karaku9/Traffic-Flow-Predictor}} for spatially informed base station planning and operation.

\end{abstract}

\ccsdesc[500]{Networks~Network simulations}
\ccsdesc[500]{Computing methodologies~Modeling and simulation}
\ccsdesc[500]{Information systems~Spatial-temporal systems}

% %%
% %% Keywords. The author(s) should pick words that accurately describe
% %% the work being presented. Separate the keywords with commas.
\keywords{Cellular traffic, Generative AI, Earth Observation}

% information and builds the first part of the formatted document.

\maketitle
% \renewcommand{\shortauthors}{Shiyuan Zhang et al.}

% \begingroup\renewcommand\thefootnote{\textsection}
% % \footnotetext{Corresponding Author.}
% \endgroup

\section{Introduction}
Fifth generation (5G) cellular networks are expected to support emerging applications such as autonomous driving, industrial automation, and smart city services~\cite{lykakis2025data, rafique2024survey}. These applications impose strict requirements on network capacity and coverage reliability in highly 
urban environments with dynamic traffic demand and complex propagation conditions~\cite{carrese2017dynamic, ghosh2024sparc, xu2025poster}. Meeting these requirements at scale requires a city-level understanding of network behavior and urban dynamics, where base station (BS) deployment and operation interact with spatial structure, traffic patterns, and environmental factors~\cite{zhang2023deep}.

\begin{figure}[t]
\centering
\includegraphics[width=0.95\linewidth]{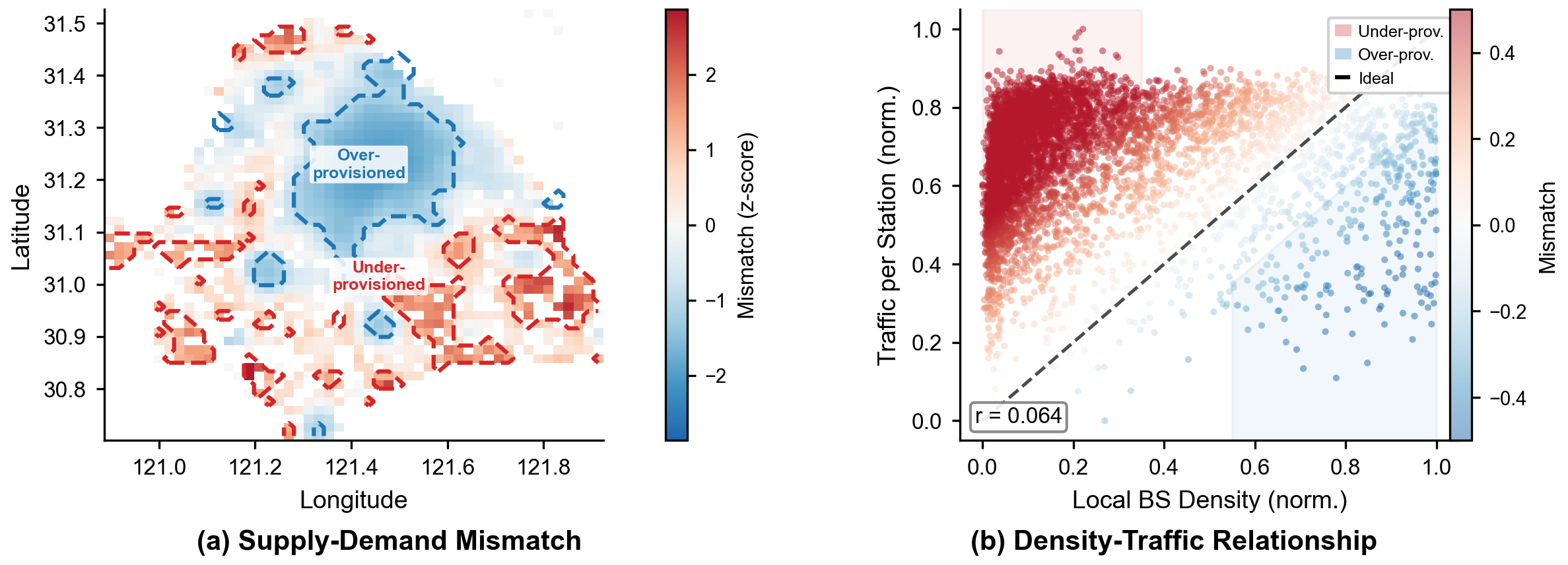}
\caption{BS density and traffic volume in Shanghai. Left: Spatial distribution of BSs shows dense deployment in urban centers, while suburban and newly developed areas lack coverage. Right: Traffic utilization analysis reveals significant capacity waste in densely deployed regions, where many BSs operate far below their capacity.}
\label{fig:Traffic_Mismatch}
\end{figure}
Despite the rapid deployment of 5G networks, managing city-scale cellular systems remains challenging due to the complex interaction between spatial deployment and dynamic traffic demand.
To ground the discussion in real-world evidence, we analyze 5{,}326 BSs across Shanghai~\cite{zhang2023deep}, one of the largest and most densely populated metropolitan areas, as shown in \figurename~\ref{fig:Traffic_Mismatch}.
The figure reveals two interrelated phenomena. First, spatial deployment is highly uneven. BSs are concentrated in established urban centers, while suburban and newly developed areas face coverage gaps. In practice, limited traffic visibility in emerging regions forces operators to rely on heuristics that favor established commercial districts, leaving fast-growing regions underserved. Second, even within densely deployed regions, substantial capacity waste persists. A significant fraction of stations operate far below their designed capacity for extended periods, while others experience periodic overload. This spatial and temporal mismatch stems from the lack of real-time traffic prediction, which could enable dynamic load balancing and energy-saving sleep scheduling, resulting in static operation, avoidable energy waste, and inefficient resource utilization.

% The major approaches to cellular network planning traditionally rely on expert-driven site selection and radio propagation simulations, which are effective for localized optimization but difficult to scale to city-wide deployments and often fail to account for fine-grained, heterogeneous traffic demand.
Traditional cellular network planning relies on expert-driven site selection and radio-propagation simulations~\cite{hui2023large, lee2014spatial, chi2024rf, erni2022adaptover, zhang2020microscope}. These approaches are effective for localized coverage optimization but difficult to scale to city-wide deployments and often fail to capture fine-grained demand heterogeneity.
To enhance planning efficiency, data-driven demand modeling has been introduced, using historical traffic patterns, demographic data, points of interest (POIs), and mobility traces to estimate or forecast traffic demand~\cite{chai2025uomo, yuan2024unist}.
However, existing approaches depend heavily on abundant historical traffic measurements, making them inapplicable to newly developed areas without prior data~\cite{xu2021spectragan}.
Furthermore, even for existing networks, autoregressive forecasting models accumulate errors over long horizons and struggle to capture stochastic aperiodic fluctuations, limiting their utility for long-term planning and real-time operation~\cite{xu2016short, sun2024netgsr}.

A practical opportunity lies in the rich spatial context that urban environments inherently contain.
Satellite imagery and POI data collectively encode urban semantics from physical infrastructure to functional activity patterns that strongly correlate with communication demand~\cite{zhang2025lsdm, xiao2025llm, sun2025deepspace, liu2025poster, hassan2024spaceris}. Motivated by this observation, we aim to learn a direct mapping from the multi-modal urban context to traffic demand. Such a mapping enables demand-aware deployment in regions where historical traffic measurements are unavailable, while also supporting accurate traffic prediction for existing networks. This unified modeling system enables long-term planning and real-time operation within a consistent demand modeling paradigm, supporting decisions such as demand-aware site selection, dynamic sleep scheduling, and load balancing.
Realizing this city-level demand intelligence requires addressing several technical challenges.

$\bullet$ \textbf{Heterogeneous Context Representation.}
Urban demand is shaped by diverse signals varying in modality, resolution, and noise.
Effectively leveraging such data requires aligning heterogeneous signals across geographic space and modeling their cross-modal interactions that predict traffic demand.
The learned representation must remain robust to missing, sparse, or imperfect modalities that commonly arise, while capturing stable, transferable patterns that generalize across different urban environments.

$\bullet$ \textbf{Long-Horizon Demand Synthesis Under Uncertainty.}
BS deployment is inherently a long-term decision depending on realistic demand evolution.
This requires synthesizing demand trajectories that preserve both periodic structure and stochastic aperiodic variations, while avoiding error accumulation that commonly affects autoregressive models~\cite{xu2014network, xu2023hybrid}.
The synthesized demand should capture key spatio-temporal statistics and support reliable long-horizon what-if analysis.

$\bullet$ \textbf{Real-Time Operational Optimization.}
% Beyond long-term planning, operators require real-time strategies for existing networks, including dynamic sleep scheduling and load balancing.
% This necessitates a fast inference engine capable of accurate traffic prediction under tight latency constraints, enabling what-to-do decision-making that reduces energy costs while preserving quality of service (QoE).
Beyond long-term planning, operators require real-time strategies for existing networks. The inference engine must produce accurate traffic predictions within millisecond-level latency while remaining consistent with the long-horizon model used for planning, so that operational what-to-do decisions align with deployment assumptions.

We introduce NetSpatial, a unified system that addresses both network planning and operation through multimodal spatial context-conditional traffic generation.
NetSpatial learns a multimodal encoder to align heterogeneous spatial signals, capturing the complex association between urban semantics and communication demand.
Employing a multi-level flow matching architecture, NetSpatial generates complete 672-hour traffic sequences directly, separately modeling periodic trends and stochastic residuals to mitigate error compounding while preserving realistic variability.
The same generative framework supports dual decision scenarios: what-if analysis ranks candidate sites based on deployment‑aware utility to support deployment planning, and what-to-do decisions use real-time traffic forecasts to guide BS sleep scheduling and load balancing for operational optimization.
The contributions of our work can be summarized as follows.

$\bullet$ We design a multimodal spatial context alignment module that integrates physical environment and functional semantics into a unified spatial representation. Specifically, satellite imagery captures the physical urban structure while POI distributions encode functional activity patterns. By aligning these heterogeneous signals into a shared representation, the module establishes transferable associations between urban semantics and cellular demand.

$\bullet$ We propose a multi-level flow matching generative architecture for long-horizon traffic synthesis that effectively reduces error accumulation.

% $\bullet$ We demonstrate that the unified generative framework supports both what-if analysis for deployment planning and what-to-do decisions for real-time operation, bridging long-term infrastructure planning with short-term network management.

$\bullet$ We show that the same generative backbone supports both deployment utility-aware site ranking for deployment planning and real-time traffic forecasting for sleep scheduling and load balancing, with no architectural modification between the two modes.

% $\bullet$ We evaluate NetSpatial on real-world dataset, demonstrating superior performance over baselines in both cross-region traffic prediction and capacity-aware site selection tasks.

$\bullet$ We evaluate NetSpatial on real-world datasets. The proposed system reduces JSD by over 29.44\% compared with the strongest baseline, generalizes across cities in zero-shot experiments, and enables up to 16.8\% energy reduction while maintaining over 80\% quality of experience (QoE) in downstream operational simulations. The system has been deployed as an online platform for BS planning.

\section{Preliminaries}

\subsection{Traffic Signal Decomposition}
Cellular traffic exhibits strong temporal regularities driven by human activity, with clear daily and weekly patterns widely observed in mobile network measurements~\cite{hui2023large}.
Let $x(t)$ denote the traffic volume of a BS at time $t$.
We decompose the signal into a structured periodic component and a residual component as
\begin{equation}
x(t) = p(t) + r(t),
\label{eq:traffic_decomposition}
\end{equation}
where $p(t)$ captures regular temporal patterns and $r(t)$ captures non-periodic fluctuations.
\begin{figure*}[!t]
\centering
\includegraphics[width=0.95\linewidth]{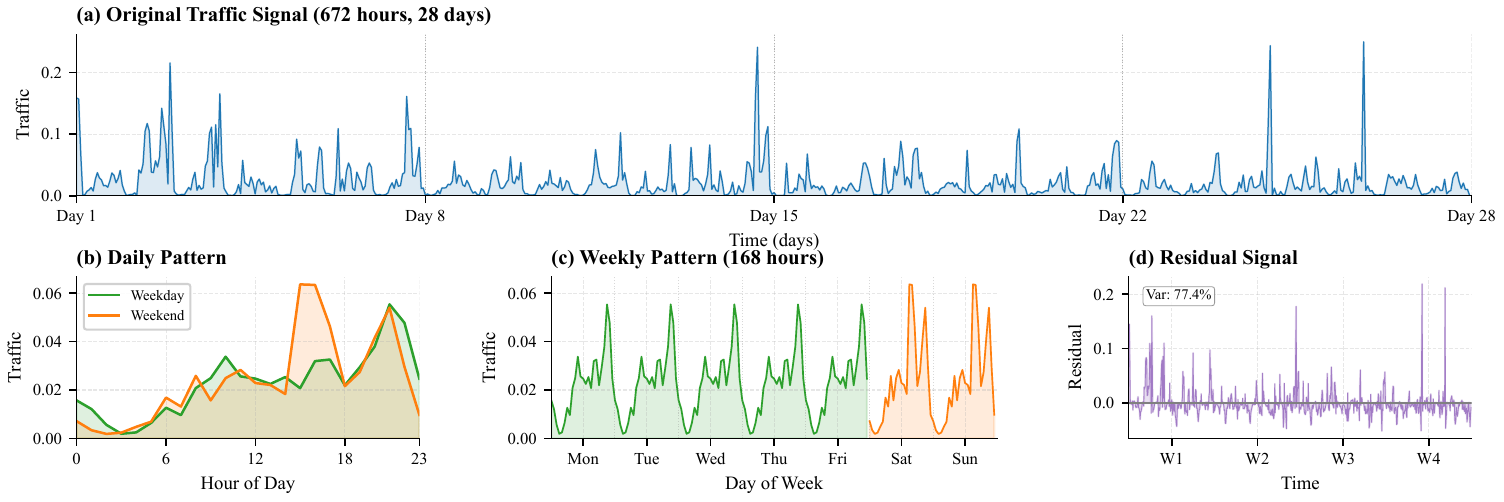}
\caption{Decomposition of BS traffic into periodic and non-periodic components. (a) Original 28-day traffic signal at hourly granularity. (b) Daily patterns for weekdays and weekends. (c) Weekly template over 168 hours. (d) Residual signal captures non-periodic fluctuations.}
\Description{Plots showing decomposition of base station traffic: original signal, daily and weekly patterns, and residuals.}
\label{traffic_dec}
\end{figure*}
The periodic component contains regularities at multiple time scales.
As shown in Fig.~\ref{traffic_dec}(b) and (c), we first consider a daily pattern over hours of the day.
Let $\tau \in [0,24)$ denote the hour within a day.
To reflect behavioral differences between weekdays and weekends, we introduce two daily patterns, i.e., a weekday pattern $d_w(\tau)$ and a weekend pattern $d_e(\tau)$,
which patterns typically show daytime peaks and nighttime troughs.

We then build a comprehensive weekly template spanning 168 hours by combining five weekday patterns followed by two weekend patterns,
\begin{equation}
w(t) =
\begin{cases}
d_w(t \bmod 24), & \text{if } \left\lfloor \frac{t}{24} \right\rfloor \bmod 7 < 5, \\
d_e(t \bmod 24), & \text{otherwise}.
\end{cases}
\label{eq:weekly_pattern}
\end{equation}

We obtain the periodic component directly by repeating the weekly template,
\begin{equation}
p(t) = w(t \bmod 168).
\label{eq:periodic_component}
\end{equation}
The residual component is precisely the remainder after removing periodicity,
\begin{equation}
r(t) = x(t) - p(t),
\label{eq:residual_component}
\end{equation}
which captures non-periodic fluctuations arising from stochastic events, anomalies, and short-term variations that are not explained by regular temporal structures.
As shown in \figurename~\ref{traffic_dec}, this decomposition effectively  separates deterministic periodic behavior from residual dynamics, which motivates a multi-level modeling approach that treats periodic structure and residual uncertainty differently in subsequent traffic generation.

\subsection{Spatial and Temporal Characteristics of Cellular Traffic}
Cellular traffic patterns are strongly influenced by the surrounding urban environment~\cite{zhang2023deep, hui2023large}.
BSs located in different functional regions, such as office, commercial, residential, and entertainment areas, exhibit distinctive temporal usage patterns, while BSs in similar environments tend to show comparable traffic patterns.

\begin{figure*}[t]
\centering
\includegraphics[width=0.95\linewidth]{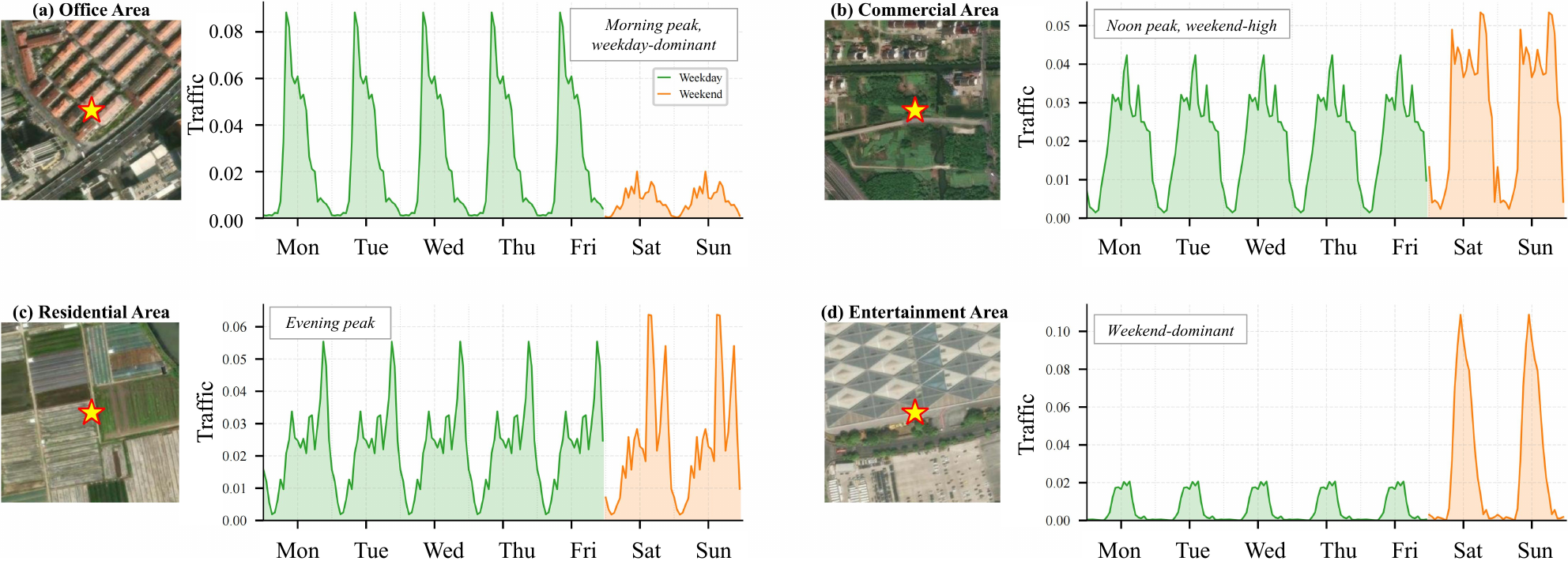}
\caption{Spatial influence on BS traffic patterns. BSs from different functional areas exhibit distinct weekly profiles. Office areas show weekday-dominant daytime peaks, commercial areas show noon peaks with higher weekend activity, residential areas show evening peaks, and entertainment areas show weekend-dominant traffic.}
\label{spatial_traffic}
\end{figure*}

The physical environment in the city shapes how traffic demand evolves over time. To illustrate this relationship, we examine representative BSs located in different functional areas of Shanghai. As shown in \figurename~\ref{spatial_traffic}, clear differences appear in peak timing and weekday--weekend contrast across these regions.
These observations indicate that urban semantics is a key driver of spatial heterogeneity in traffic demand.

Beyond the physical environment, the functional semantics of an area also affect traffic variability. 
% As shown in \figurename~\ref{spatial_diff}, 
Human activity tends to follow more regular patterns in central districts, leading to relatively stable traffic demand.
In outer districts, demand varies more sharply, with larger fluctuations.
These differences arise from the types of activities occurring in each area, as reflected in the distribution of nearby POIs.

% \begin{figure*}[t]
% \centering
% \includegraphics[width=0.95\linewidth]{figs/rs2.pdf}
% \caption{The central panel presents a heatmap of traffic fluctuation across BSs in the city.
% The residual variance ratio
% $R_i = \sigma^2_{\mathrm{res},i} / \sigma^2_{\mathrm{total},i}$
% measures the fraction of traffic variance unexplained by periodic
% patterns. Lower values (green) indicate regular, predictable demand,
% while higher values (red) indicate irregular, stochastic demand.}
% \label{spatial_diff}
% \end{figure*}

Formally, let $\mathbf{c}_i = [\mathbf{e}_i;\, \mathbf{z}_i]$ denote the spatial context at location $i$, where $\mathbf{e}_i$ represents the physical environment features extracted from satellite imagery, and $\mathbf{z}_i$ represents the functional semantics derived from the surrounding POI distribution.
We express the site traffic at location $i$ as
% \begin{equation}
% x_i(t) = p_i(t) + r_i(t)=f_1(\mathbf{c}_i)+f_2(\mathbf{c}_i)
% \end{equation}
\begin{equation}
\mathbf{x}_i^{1:T} = \mathbf{p}_i^{1:T} + \mathbf{r}_i^{1:T} = f_1(\mathbf{c}_i) + f_2(\mathbf{c}_i)
\end{equation}
where $\mathbf{x}_i^{1:T} = \{x_i(t)\}_{t=1}^{T}$ represents the complete time series for base station $i$ over $T$ time steps, $\mathbf{p}_i^{1:T}$ and $\mathbf{r}_i^{1:T}$ represent the predictable and residual components respectively, and $f_1(\mathbf{c}_i)$ and $f_2(\mathbf{c}_i)$ are functions conditioned on the spatial context $\mathbf{c}_i$.
\subsection{From Measured BSs to Feasible Candidate Sites}

Our training data provides traffic traces only at deployed base stations,
while deployment planning requires evaluating many candidate locations.
We bridge this gap by modeling traffic as a location-conditional
site-traffic potential function.
During training, each deployed BS provides a labeled sample consisting of the surrounding spatial context $\mathbf{c}$ as input and the measured traffic time series as the corresponding label. The model learns a conditional generator
$G_\theta\!: \mathbf{c}_i \mapsto P\!\left(\mathbf{x}_i \mid \mathbf{c}_i\right)$
that maps spatial context to a distribution over traffic sequences.
During inference, let $\mathcal{C}=\{1,\dots,M\}$ denote the set of $M$ feasible grid cells that satisfy basic engineering constraints.
For each cell~$j\in\mathcal{C}$, we extract its spatial
context~$\mathbf{s}_j$ and generate a monthly traffic sequence
$\hat{\mathbf{x}}_j=[\hat{x}_j(1),\dots,\hat{x}_j(T)]$, interpreted as the traffic a BS would carry if deployed at cell~$j$.
Because the generator produces a complete traffic sequence for each candidate location, the downstream site-ranking objective can be defined flexibly to suit different planning goals. For example, operators may prioritize locations with high demand, high peak load, or potential revenue. As an illustrative example, we introduce a simple Load Stability Index (LSI) that favors locations with relatively stable traffic patterns over time. Such stability is desirable in practical deployment scenarios where operators prefer locations with consistently high traffic volume rather than highly volatile demand that may lead to frequent over-provisioning or congestion. Formally, we define the LSI as
\begin{equation}\label{eq:lsi}
  \mathrm{LSI}_j
  = \frac{1}{\,\sigma(\hat{\mathbf{x}}_j)+\epsilon\,},
\end{equation}
where $\sigma(\hat{\mathbf{x}}_j)$ is the temporal standard deviation of
the generated sequence and $\epsilon=10^{-6}$.
Given a budget of $K$ new sites, we select
\begin{equation}\label{eq:topk}
  \mathcal{S}^*
  = \underset{\mathcal{S}\subseteq\mathcal{C},\;|\mathcal{S}|=K}
    {\arg\max}\;\sum_{j\in\mathcal{S}}\mathrm{LSI}_j\,,
\end{equation}
which reduces to sorting by LSI and retaining the top~$K$ entries due to the objective's modularity.

Importantly, the traffic generator and the ranking objective are decoupled. Since the model produces a full traffic sequence for each candidate location, alternative deployment utilities can be applied directly without retraining.

\section{System Overview}
\label{sec:system}
This section presents NetSpatial, a unified system for cellular network planning and operation based on spatially conditional traffic generation.
The core of NetSpatial is a multi-level conditional generative model that fuses satellite imagery and POI distributions to map spatial context to long-term site traffic sequences, enabling both what-if analysis for BS deployment planning and what-to-do decisions for real-time network operation.

\subsection{Problem Setting and Outputs}
\label{sec:system_setting}

For BSs with traffic measurements, each station forms a labeled training sample. The label is an hourly traffic sequence over a 28-day window with a length of 672. The input is the spatial multimodal context around the station, derived from satellite imagery and POI distributions. The model learns to generate a 672-hour site traffic sequence conditioned on this spatial context.

\textbf{Inference on candidate locations.}
For candidate locations that lack traffic data, namely potential deployment sites, NetSpatial evaluates a set of feasible candidates derived from grid cells after applying deployability constraints.
For each feasible candidate location, NetSpatial extracts spatial multi-modal context using the same pipeline as in training and generates a 672-hour hourly traffic sequence.
We interpret this sequence as the traffic load that a base station would carry if deployed at that candidate location.

\textbf{Real-time inference on existing BSs.}
For BSs already in operation, the generative model performs real-time traffic prediction using satellite imagery and POI data to capture spatial context, and periodic patterns derived from Equations~\eqref{eq:weekly_pattern} and~\eqref{eq:periodic_component} to capture temporal dynamics.
These predictions enable dynamic operational decisions, including BS sleep scheduling for energy saving during low-traffic periods and load balancing across neighboring stations to optimize network resource utilization.

\textbf{Decision output.}
NetSpatial delivers two distinct categories of decision outputs,

% $\bullet$ For network planning, it assigns each candidate site a capacity-aware utility score computed from its generated hourly traffic sequence under a constant hourly capacity $C$, and outputs the top-$K$ candidate sites by ranking these scores.
% $\bullet$ For network planning, it computes a Load Stability Index (LSI) for each candidate site based on its generated monthly traffic sequence, ranks the sites by this index, and selects the top-$K$ optimal locations.

% $\bullet$ For network operation, it provides real-time traffic forecasts that inform BS on/off switching and inter-station load redistribution decisions.
$\bullet$ For network planning, it computes LSI for each candidate site based on its generated monthly traffic sequence, ranks all the sites by this index, and then finally selects the top-$K$ optimal locations.

$\bullet$ For network operations, it provides accurate, real-time traffic forecasts that inform dynamic BS on-off switching and efficient inter-station load redistribution decisions.

\subsection{Design Overview}
\label{sec:design_overview}

\begin{figure}[t]
\centering
\includegraphics[width=1\linewidth]{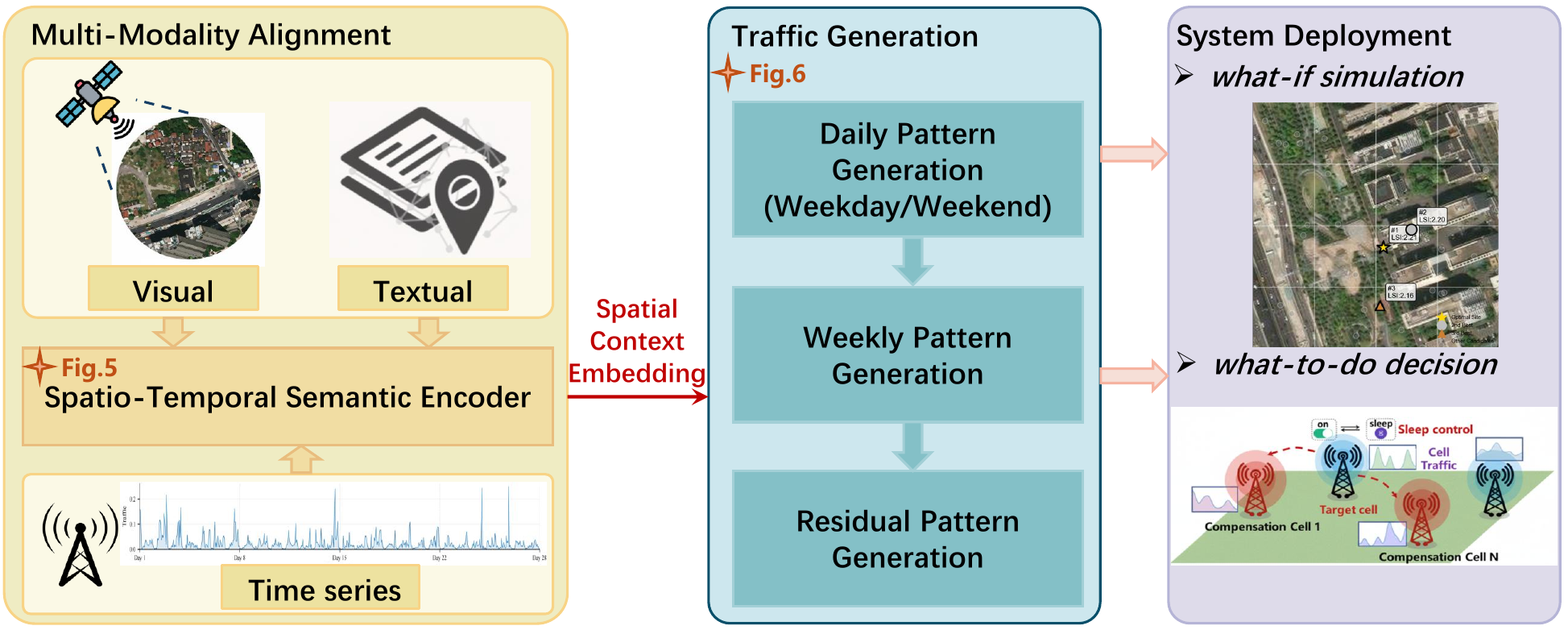}
\caption{System overview of NetSpatial.}
\label{fig:system_overview}
\end{figure}

Fig.~\ref{fig:system_overview} provides an overview of the end-to-end workflow of NetSpatial, which comprises three modules: multimodal data alignment, multi-level conditional generation, and decision support for planning and operation.
The first module ingests heterogeneous spatial data sources, including satellite imagery and POI distributions, and aligns them into a unified spatial representation using a multimodal encoder.
The second module takes the aligned multimodal context as input. It generates a 672-hour traffic sequence $\hat{x}_j(t)$ for $t \in \{1,\dots,672\}$ via a multi-level flow matching architecture that separately models periodic trends and stochastic residuals.
The third module supports two decision scenarios. In what-if analysis for network planning, candidate sites are ranked by the Load Stability Index (Eq.~\eqref{eq:lsi}) computed from their generated traffic profiles, and the top-$K$ locations with the most temporally stable demand are selected for deployment.
In what-to-do decisions for network operation, the same generator provides real-time traffic forecasts to inform BS sleep scheduling and load balancing.

\section{Methodology}
\label{sec:method}

\begin{figure}[t]
\centering
\includegraphics[width=0.95\linewidth]{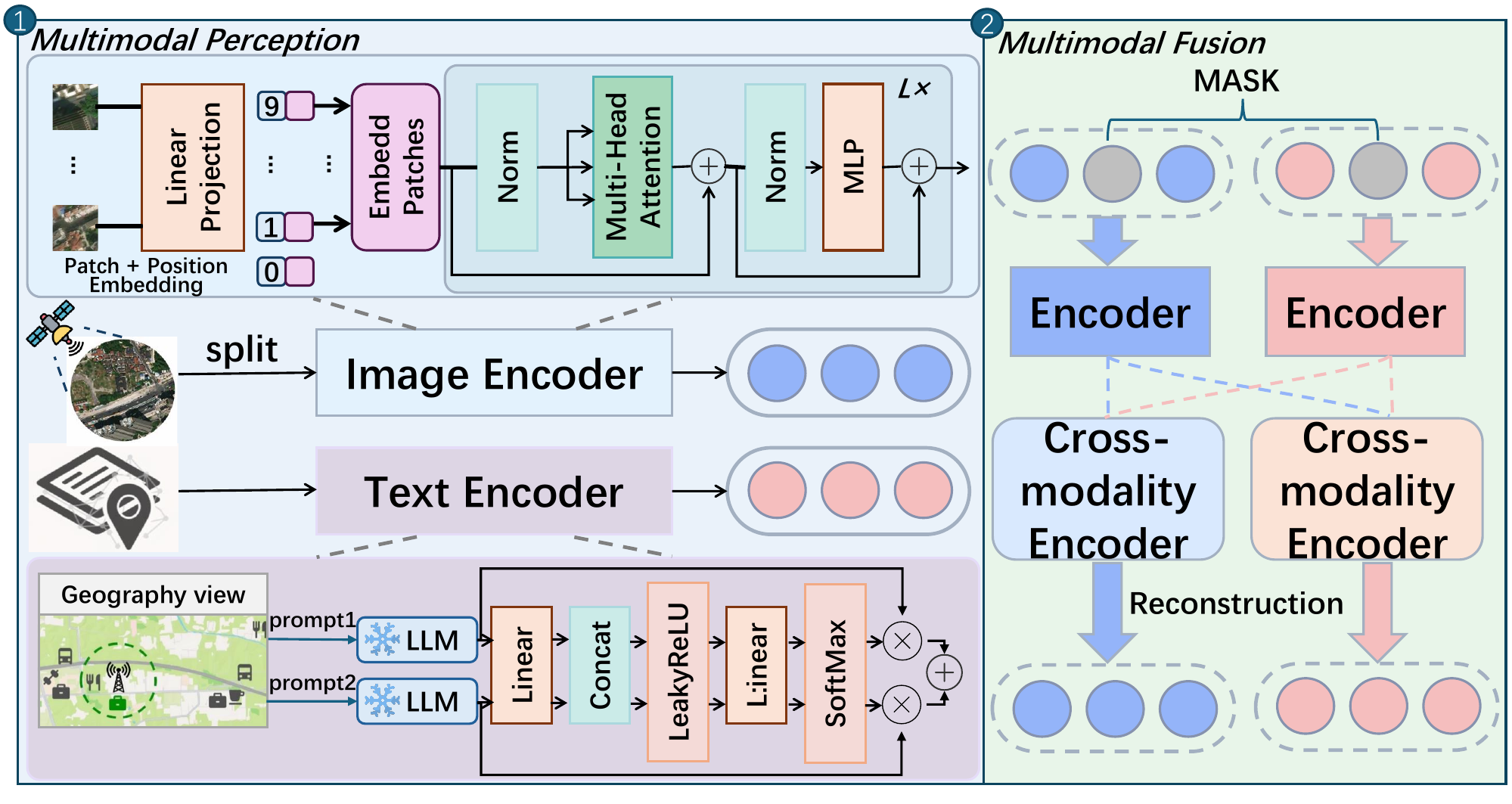}
\caption{Overview of the proposed multi-modal perception and fusion framework. It consists of two stages. (1) \textbf{Multi-modal Perception}. Dedicated frozen encoders process satellite imagery and POI data to extract modality-specific features. (2) \textbf{Multi-modal Fusion}. A mask-based generative mechanism integrates the features into a unified representation that is robust to missing modalities.}
\label{multi}
\end{figure}

\subsection{Multi-Modal Representation Learning}

As shown in Fig.~\ref{multi}, we propose a multi-modal framework that produces a unified spatial embedding $\mathbf{C}\in\mathbb{R}^d$ for each location~$\ell$.  The input consists of a geo-referenced satellite image patch $\mathbf{x}^I\in\mathbb{R}^{H\times W\times 3}$ and a set of nearby POIs $\mathcal{P}(\ell)=\{p_i\}_{i=1}^K$ within a fixed radius ($K$ may be zero).  The framework captures dense visual cues from imagery and sparse symbolic semantics from POIs while handling variable POI density, including complete absence.

\subsubsection{Multi-modal Perception}

We employ frozen pretrained encoders for stable and transferable feature extraction.

\textbf{Visual modality.}
A ViT-based encoder $f_I$ processes the satellite patch and produces a visual embedding $\mathbf{v}\in\mathbb{R}^d$ (followed by $\ell_2$-normalization, all embeddings presented below are likewise normalized before being used).

\textbf{Textual modality.}
For each POI $p_i$ we construct an address prompt $T_i^A$ and a surrounding-context prompt $T_i^S$.  A frozen LLM llama-3.1-8b~\cite{weerawardhena2025llama} $H$ extracts embeddings $\mathbf{e}_i^A, \mathbf{e}_i^S\in\mathbb{R}^D$, which are fused by a lightweight scoring network,
\begin{equation}
  [\alpha_i^A,\,\alpha_i^S]
  = \mathrm{softmax}\!\big(s_\theta(\mathbf{e}_i^A \| \mathbf{e}_i^S)\big),
  \quad
  \mathbf{e}_i = \alpha_i^A\,\mathbf{e}_i^A + \alpha_i^S\,\mathbf{e}_i^S,
  \label{eq:poi_fuse}
\end{equation}
where $\|$ denotes concatenation.  Each fused embedding is linearly projected into the shared $d$-dimensional space to obtain $\mathbf{p}_i$.

To handle variable $K$, we aggregate $\{\mathbf{p}_i\}_{i=1}^K$ directly via attention pooling,
\begin{equation}
  \beta_i = \mathrm{softmax}_i\!\big(a_\varphi(\mathbf{p}_i)\big),
  \quad
  \mathbf{p} = \textstyle\sum_{i=1}^K \beta_i\,\mathbf{p}_i
  \in\mathbb{R}^d.
  \label{eq:poi_agg}
\end{equation}
When $K=0$, the POI modality is treated as missing.

\subsubsection{Multi-modal Fusion}

We fuse $\mathbf{v}$ and $\mathbf{p}$ with a generative mask-modeling strategy that encourages cross-modal interaction and robustness to missing modalities.  Two learnable mask tokens $\mathbf{v}_{\mathrm{mask}},\mathbf{p}_{\mathrm{mask}}\in\mathbb{R}^d$ and a fusion token $\mathbf{t}_F\in\mathbb{R}^d$ are introduced.

During training we sample a binary mask $(m^I,m^P)$ with $m^I+m^P=1$ and consistently replace the masked modality with its mask token,
\begin{equation}
  \tilde{\mathbf{v}} = (1-m^I)\,\mathbf{v} + m^I\,\mathbf{v}_{\mathrm{mask}},
  \quad
  \tilde{\mathbf{p}} = (1-m^P)\,\mathbf{p} + m^P\,\mathbf{p}_{\mathrm{mask}}.
  \label{eq:mm_mask}
\end{equation}
This constraint keeps only one modality visible and the other masked at each training step, forcing the model to reconstruct the masked modality from the visible one.

A lightweight Transformer $\mathcal{T}_\theta$ processes the three-token sequence $[\mathbf{t}_F;\,\tilde{\mathbf{v}};\,\tilde{\mathbf{p}}]$ and returns hidden states $[\mathbf{h}_F,\mathbf{h}_v,\mathbf{h}_p]$.  The unified spatial embedding is read from the fusion token,
\begin{equation}
  \mathbf{z} = \mathrm{norm}(\mathbf{W}_z\,\mathbf{h}_F)
  \in\mathbb{R}^d.
  \label{eq:mm_z}
\end{equation}

Cross-modal alignment is enforced by reconstructing the masked modality from its corresponding hidden state via linear heads.  The training objective maximizes cosine similarity between the reconstruction and the original as,
\begin{equation}
  \mathcal{L}_{\mathrm{gen}}
  = \mathbb{E}_{m^I,m^P}\!\Big[
      m^I\,(1-\hat{\mathbf{v}}^\top\mathbf{v})
    + m^P\,(1-\hat{\mathbf{p}}^\top\mathbf{p})
    \Big].
  \label{eq:mm_loss}
\end{equation}

In the inference phase, when POIs are absent ($K=0$), we set $m^P=1$, allowing the model to degrade to visual-only semantics gracefully.  Only the fusion module parameters are trained, encoders $f_I$ and $H$ remain frozen.

\subsection{Multi-level Conditional Model for Monthly Traffic Generation}

\begin{figure}[t]
\centering
\includegraphics[width=0.95\linewidth]{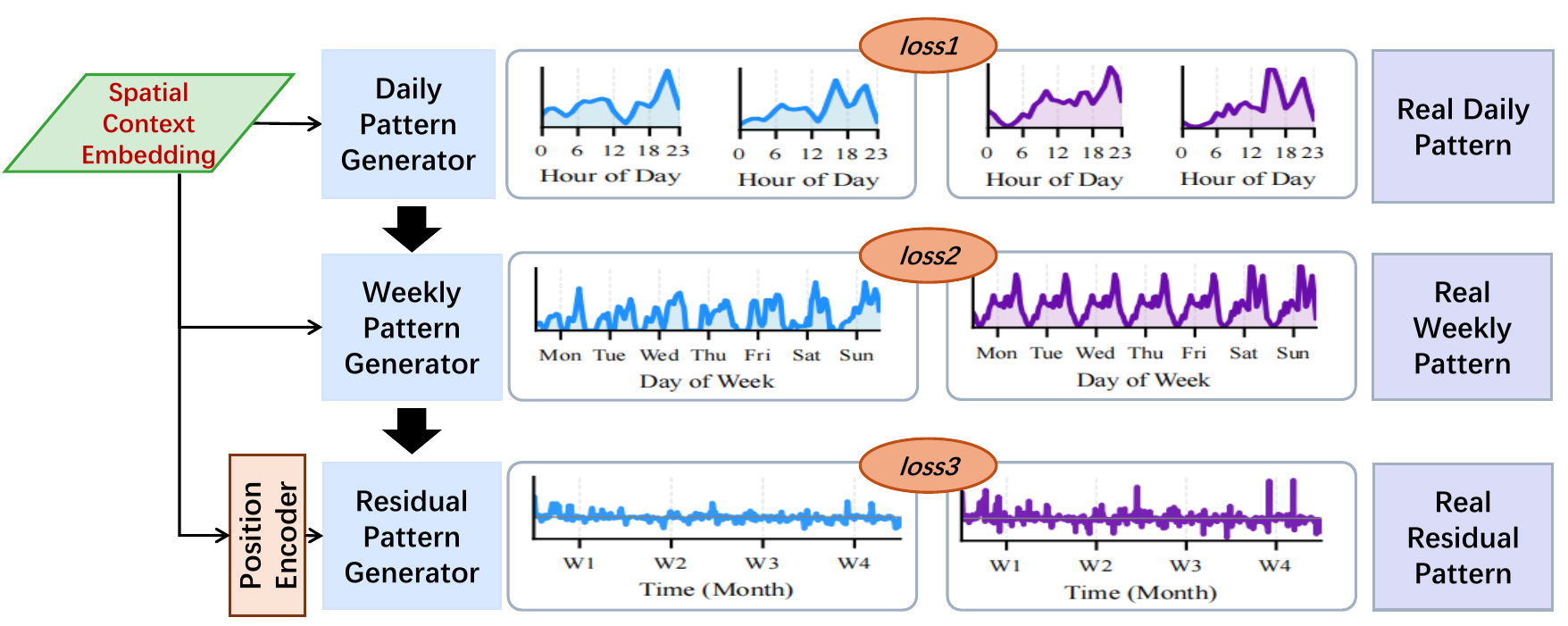}
\caption{Architecture of the multi-level conditional model for monthly traffic generation. Guided by the context embedding, the framework progressively generates the complete monthly-period traffic through hierarchical refinement.}
\label{overview}
\end{figure}

The multimodal perception and fusion module produces a shared spatial representation $C \in \mathbb{R}^d$. We use $C$ as a cross-level spatial prior and obtain level-specific conditions $c_1,c_2,c_3$ via lightweight multiscale projections.
A peak hour classifier built on $C$ produces a distribution over 24 hours and identifies the peak index $h^\star$.  This module corresponds to the \textbf{Position Encoder} in Figure~\ref{overview}, it locates where major traffic peaks and valleys occur over the monthly horizon. 
The resulting $h^\star$ is injected as an explicit condition into the residual level, anchoring the generated long-term pattern to the correct peak placement.
The generator decomposes a monthly sequence of length $T=672$ with $S=24$ hours per day into a daily template $d$, a weekly pattern $w$, a periodic component $u$, and a residual $r$
\begin{equation}
\begin{aligned}
d &= G_1(z_1,c_1), \quad
w = G_2(z_2,d,c_2), \quad
u = \operatorname{Rep}(w), \\
r &= G_3(z_3,u,d,w,c_3,h^\star), \quad
\hat x = \operatorname{clippos}(u + r) .
\end{aligned}
\label{eqgen}
\end{equation}
Here $z_1,z_2,z_3\sim\mathcal{N}(0,I)$ are level specific noise inputs, $G_1,G_2,G_3$ are the conditional generators for the daily template, weekly pattern, and residual, respectively, $\operatorname{Rep}$ repeats $w$ to length $T$, and $\operatorname{clippos}(x)=\max(x,0)$ enforces nonnegative traffic. Each level is trained with flow matching to regress the target velocity field
\begin{equation}
\mathcal{L}_{fm}^\ell = \mathbb{E}\left\| v_\theta^\ell(x_t,t,c_\ell,\cdot) - v^\star \right\|_2^2 .
\label{eqfm}
\end{equation}
We sample $x_0\sim\mathcal{N}(0,I)$ and $t\sim\mathcal{U}(0,1)$, set $x_t=t x_1+(1-t)x_0$ and $v^\star=x_1-x_0$, and $v_\theta^\ell$ denotes the level $\ell$ velocity field. Auxiliary losses enforce nonnegativity, temporal smoothness, daily and weekly periodic consistency, mean bias correction, peak hour classification, and correlation alignment. The objective is the weighted sum
\begin{equation}
\label{eqobj}
\mathcal{L} = \sum_{\ell=1}^{3} \lambda_\ell \, \mathcal{L}_{fm}^{\ell} + \sum_{p \in \mathcal{P}} \lambda_p \, \mathcal{L}_p
\end{equation}
where $\mathcal{P}=\{\text{bnd},\text{tmp},\text{per},\text{bias},\text{peak},\text{corr}\}$, $\mathcal{L}_p$ denotes the corresponding auxiliary loss~\cite{li2024mobile}, and all $\lambda_\ast$ are strictly nonnegative weights.

\subsection{Generator Algorithm and Inference Efficiency}

% In Algorithms~\ref{alg:train} and \ref{alg:infer}, the shared spatial representation is $C \in \mathbb{R}^d$ and the peak classifier uses a weight matrix $W_C$ to produce $q=\operatorname{softmax}(W_C C)$ with peak index $h^\star=\arg\max_h q_h$. The real sequence is $x \in \mathbb{R}^T$ with $T=672$, $S=24$, $D=28$, and $W=4$. Reshaping yields $X \in \mathbb{R}^{D\times S}$. The weekday and weekend means form the daily target $d_{tar}=[d_{wk}\;d_{we}] \in \mathbb{R}^{2S}$, and the weekly target $w_{tar}\in\mathbb{R}^{7S}$ is obtained by averaging the same weekday across the $W$ weeks. The periodic target is $u_{tar}=\operatorname{Rep}(w_{tar})$ and the residual target is $r_{tar}=x-u_{tar}$. The multiscale projections map $C$ to $c_1,c_2,c_3$. The operators $\operatorname{Rep}$, $\operatorname{clipsym}(\cdot,b)$, and $\operatorname{clippos}(\cdot)$ denote weekly repetition, symmetric clipping with bound $b$, and nonnegative clipping, with bounds $b_1,b_2,b_3$ for the three levels. The step counts are $N_1,N_2,N_3$ with step sizes $\Delta t_i=1/N_i$. Level three uses scaled noise $x\sim 0.1\,\mathcal{N}(0,1)$ as in the implementation.
Algorithms~\ref{alg:train} and~\ref{alg:infer} detail the training and inference procedures.  Both operate on a shared spatial embedding $C\in\mathbb{R}^d$, from which level-specific conditions $c_1,c_2,c_3$ are derived via learned projections.  A linear peak classifier produces $q=\operatorname{softmax}(W_C C)$ and extracts the peak hour index $h^\star=\arg\max_h q_h$, which anchors the residual level to the correct diurnal peak.

\paragraph{Hierarchical target decomposition.}
Given a real traffic sequence $x\in\mathbb{R}^T$ spanning $W{=}4$ weeks ($T{=}D\!\times\!S$, $D{=}28$ days, $S{=}24$ hours), we decompose it into three targets of increasing temporal resolution.
\begin{enumerate}[nosep,leftmargin=1.5em]
  \item \emph{Daily target}
        $d_{\mathrm{tar}}=[d_{\mathrm{wk}}\;d_{\mathrm{we}}]
        \in\mathbb{R}^{2S}$, the average hourly profile for
        weekdays and weekends respectively.
  \item \emph{Weekly target}
        $w_{\mathrm{tar}}\in\mathbb{R}^{7S}$, the average hourly
        profile for each day of the week, computed across the $W$ weeks.
  \item \emph{Residual target}
        $r_{\mathrm{tar}}=x-u_{\mathrm{tar}}\in\mathbb{R}^T$, the
        deviation from the periodic component
        $u_{\mathrm{tar}}=\operatorname{Rep}(w_{\mathrm{tar}})$, where
        $\operatorname{Rep}(\cdot)$ tiles a weekly pattern to length~$T$.
\end{enumerate}
Each level learns to generate a target conditional on the outputs of all coarser levels, preventing errors from accumulating across scales.

\paragraph{Generation and post-processing.}
All three levels share the same Euler integration scheme with $N_\ell$ steps and step size $\Delta t_\ell=1/N_\ell$.  After each step we apply $\operatorname{clipsym}(\cdot,b_\ell)$, which clips values to $[-b_\ell,\,b_\ell]$, to keep intermediate states within a physically plausible range.  Level~3 initializes from scaled noise $x\sim 0.1\,\mathcal{N}(0,I)$ rather than unit Gaussian, reflecting the smaller magnitude of residual fluctuations relative to the periodic component.  The final output is assembled as Equation~\eqref{eqgen}.

\begin{algorithm}[t]
\caption{Hierarchical flow-matching training}
\label{alg:train}
\begin{algorithmic}[1]
\Require real sequence $x\!\in\!\mathbb{R}^T$, spatial embedding $C$
\Ensure  total loss $\mathcal{L}$
\Statex \textit{// Shared preparation}
\State $q\!=\!\operatorname{softmax}(W_C C)$;\;
       $h^\star\!=\!\arg\max_h q_h$;\;
       $c_1,c_2,c_3\!\leftarrow\!\text{project}(C)$
\State Compute targets $d_{\mathrm{tar}},\,w_{\mathrm{tar}},\,u_{\mathrm{tar}},\,r_{\mathrm{tar}}$ from $x$
\State Sample $t\!\sim\!\mathcal{U}(0,1)$
\Statex \textit{// Level 1: daily pattern}
\State $z_1\!\sim\!\mathcal{N}(0,I)$;\;
       $x_t\!=\!t\,d_{\mathrm{tar}}+(1\!-\!t)\,z_1$;\;
       $v^\star\!=\!d_{\mathrm{tar}}-z_1$
\State $\mathcal{L}_{fm}^1\!\leftarrow\!\|G_1(x_t,t,c_1)-v^\star\|^2$
       \Comment{Eq.~\eqref{eqfm}}
\Statex \textit{// Level 2: weekly pattern}
\State $z_2\!\sim\!\mathcal{N}(0,I)$;\;
       $x_t\!=\!t\,w_{\mathrm{tar}}+(1\!-\!t)\,z_2$;\;
       $v^\star\!=\!w_{\mathrm{tar}}-z_2$
\State $\mathcal{L}_{fm}^2\!\leftarrow\!\|G_2(x_t,t,c_2,d_{\mathrm{tar}})-v^\star\|^2$
\Statex \textit{// Level 3: residual}
\State $z_3\!\sim\!\mathcal{N}(0,I)$;\;
       $x_t\!=\!t\,r_{\mathrm{tar}}+(1\!-\!t)\,z_3$;\;
       $v^\star\!=\!r_{\mathrm{tar}}-z_3$
\State $\mathcal{L}_{fm}^3\!\leftarrow\!\|G_3(x_t,t,c_3,d_{\mathrm{tar}},w_{\mathrm{tar}},u_{\mathrm{tar}},h^\star)-v^\star\|^2$
\Statex \textit{// Auxiliary losses}
\State Generate $\hat{x}$ via Algorithm~\ref{alg:infer}
\State Compute $\{\mathcal{L}_p\}_{p\in\mathcal{P}}$ (boundary, temporal, periodic, bias, peak, correlation)
\State \Return $\mathcal{L}=\sum_{\ell=1}^{3}\lambda_\ell\mathcal{L}_{fm}^\ell + \sum_{p\in\mathcal{P}}\lambda_p\mathcal{L}_p$
       \Comment{Eq.~\eqref{eqobj}}
\end{algorithmic}
\end{algorithm}

\begin{algorithm}[t]
\caption{Hierarchical flow-matching inference}
\label{alg:infer}
\begin{algorithmic}[1]
\Require spatial embedding $C$, step counts $N_1,N_2,N_3$
\Ensure  generated sequence $\hat{x}\!\in\!\mathbb{R}^T$
\State $q\!=\!\operatorname{softmax}(W_C C)$;\;
       $h^\star\!=\!\arg\max_h q_h$;\;
       $c_1,c_2,c_3\!\leftarrow\!\text{project}(C)$
\Statex \textit{// Level 1: daily pattern ($\mathbb{R}^{2S}$)}
\State $x\!\sim\!\mathcal{N}(0,I)$
\For{$k=1,\dots,N_1$}
  \State $x \leftarrow \operatorname{clipsym}\!\big(x + \tfrac{1}{N_1}\,G_1(x,\tfrac{k}{N_1},c_1),\;b_1\big)$
\EndFor
\State $d \leftarrow x$
\Statex \textit{// Level 2: weekly pattern ($\mathbb{R}^{7S}$)}
\State $x\!\sim\!\mathcal{N}(0,I)$
\For{$k=1,\dots,N_2$}
  \State $x \leftarrow \operatorname{clipsym}\!\big(x + \tfrac{1}{N_2}\,G_2(x,\tfrac{k}{N_2},c_2,d),\;b_2\big)$
\EndFor
\State $w \leftarrow x$;\; $u \leftarrow \operatorname{Rep}(w)$
\Statex \textit{// Level 3: residual ($\mathbb{R}^{T}$)}
\State $x\!\sim\!0.1\,\mathcal{N}(0,I)$
       \Comment{scaled noise}
\For{$k=1,\dots,N_3$}
  \State $x \leftarrow \operatorname{clipsym}\!\big(x + \tfrac{1}{N_3}\,G_3(x,\tfrac{k}{N_3},c_3,d,w,u,h^\star),\;b_3\big)$
\EndFor
\State $r \leftarrow x$
\State \Return $\hat{x}$
       \Comment{Eq.~\eqref{eqgen}}
\end{algorithmic}
\end{algorithm}

\section{Experiments}
\subsection{Experimental Settings}
We evaluate NetSpatial on real-world cellular traffic datasets by comparing it with state-of-the-art baselines, demonstrating both what-if and what-to-do decision scenarios, and conducting ablation studies.

\subsubsection{Dataset.} We collected a large-scale cellular traffic dataset from Shanghai, covering 5,326 BSs, gathered over one month in 2021 with hourly traffic information for each station.

\subsubsection{Baselines.} We compare our proposed model with the following six baselines. Notably, we evaluate each model with two different kinds of inputs:
1) noise vectors only, 2) noise vectors and multimodal fusion representation $C$.

\textbf{Time-series GAN}~\cite{yoon2019time}. Time-series GAN combines an LSTM in both the generator and the discriminator to generate time-series data.

\textbf{Trans-GAN}~\cite{jiang2021transgan}. TransGAN is a transformer-based generative adversarial network (GAN) that combines a multi-scale discriminator to capture low-level textures and semantic contexts with a generator using transformer blocks. We adjust the transformer block sizes for network traffic data generation in our case.

\textbf{TCN-GAN}~\cite{mogren2016continuous}. TCN-GAN is a GAN that consists of a generator and a discriminator, both of which use temporal convolutional networks (TCNs) for the generation of network traffic data.

\textbf{KeGAN}~\cite{zhang2023deep}. A hierarchical GAN that utilizes a self-constructed urban knowledge graph (UKG) to explicitly incorporate urban features during the forecasting process. 

\textbf{CSDI}~\cite{tashiro2021csdi}. A conditional diffusion model incorporating 2D attention to capture complex dependencies in structured inputs for precise conditional generation.

\textbf{DiT}~\cite{peebles2023scalable}. A novel diffusion model utilizes a transformer backbone to process latent image patches instead of the traditional U-Net architecture. 

\subsubsection{Evaluation Metrics}
We assess our model using the following set of metrics.

\textbf{Root Mean Squared Error (RMSE).} RMSE is the square root of MSE, providing an interpretable error metric in the same unit as the target variable.
\begin{equation}
\text{RMSE} = \sqrt{\frac{1}{n} \sum_{i=1}^{n} (y_i - \hat{y}_i)^2}.
\end{equation}

\textbf{Mean Absolute Error (MAE).} MAE measures the average magnitude of errors without considering their direction. A smaller value indicates better performance.
\begin{equation}
\text{MAE} = \frac{1}{n} \sum_{i=1}^{n} |y_i - \hat{y}_i|.
\end{equation}

\textbf{Jensen–Shannon Divergence (JSD).} JSD~\cite{fuglede2004jensen} is a commonly used metric to describe the similarity between two distributions, which is defined as,
\begin{equation}
\footnotesize
 JSD({\bf \hat{X}},{\bf {X}}) = \sqrt{\frac{KL({\bf {X}} \parallel {\bf \hat{X}})+KL({\bf \hat{X}}\parallel{\bf {X}})}{2}},   
\end{equation}
where ${\bf {X}}$ represents the real data, ${\bf \hat{X}}$ represents the generated data, and KL$(\cdot)$ is the Kullback-Leibler divergence. A lower JSD implies a better generation model since the distribution of the generated data is more similar to the real data.

\textbf{First-order Difference.} To evaluate the variation between every two adjacent generated traffic point, we compute the first-order difference of time series, denoted by ${\bf \hat{X}}_d = {\hat{x}}_{t+1} - {\hat{x}}_{t}$. We then compute the JSD between the first-order differences of generated datasets and their corresponding real dataset.

% 为保证相对的公平性，我们选择以相同的epoch（100）训练增加condition的相关模型，
\begin{table*}[t]
\centering
\caption{Overall comparison. $\Delta$: relative improvement of Ours over each method; $\Delta_c$: improvement from adding conditioning (+C). Lower is better for all metrics. Cell color encodes $\Delta$ magnitude: \colorbox{pos3}{\textcolor{white}{green}} = Ours is better; \colorbox{neg3}{\textcolor{white}{red}} = Ours is worse.}
\label{main_table}
\setlength{\tabcolsep}{4.5pt}
\renewcommand{\arraystretch}{1.18}
\small
\begin{tabular}{l ccc ccc ccc ccc}
\toprule
\rowcolor{gray!15}
\textbf{Method} &
\multicolumn{3}{c}{\textbf{JSD}} &
\multicolumn{3}{c}{\textbf{JSD\_DIFF}} &
\multicolumn{3}{c}{\textbf{RMSE}} &
\multicolumn{3}{c}{\textbf{MAE}} \\
\cmidrule(lr){2-4}\cmidrule(lr){5-7}\cmidrule(lr){8-10}\cmidrule(lr){11-13}
\rowcolor{gray!15}
 & Val & $\Delta$ & $\Delta_c$
 & Val & $\Delta$ & $\Delta_c$
 & Val & $\Delta$ & $\Delta_c$
 & Val & $\Delta$ & $\Delta_c$ \\
\midrule

RNN-GAN
 & 0.8326 & \dpos{93.33\%}{93} & \dpos{17.75\%}{17}
 & 0.3040 & \dpos{88.86\%}{88} & \dneg{-0.61\%}{0}
 & 0.3875 & \dpos{94.23\%}{94} & \dneg{-95.73\%}{95}
 & 0.3866 & \dpos{95.86\%}{95} & \dneg{-69.45\%}{69} \\

\quad + $C$
 & 0.6848 & \dpos{91.89\%}{91} & \dzero
 & 0.3058 & \dpos{88.93\%}{88} & \dzero
 & 0.7585 & \dpos{97.05\%}{97} & \dzero
 & 0.6552 & \dpos{97.56\%}{97} & \dzero \\
\addlinespace[2pt]

TCN-GAN
 & 0.6931 & \dpos{91.99\%}{91} & \dpos{0.00\%}{0}
 & 0.3040 & \dpos{88.86\%}{88} & \dpos{5.02\%}{5}
 & 0.4601 & \dpos{95.14\%}{95} & \dneg{-1.79\%}{1}
 & 0.4500 & \dpos{96.45\%}{96} & \dneg{-2.97\%}{2} \\

\quad + $C$
 & 0.6931 & \dpos{91.99\%}{91} & \dzero
 & 0.2887 & \dpos{88.28\%}{88} & \dzero
 & 0.4683 & \dpos{95.23\%}{95} & \dzero
 & 0.4634 & \dpos{96.55\%}{96} & \dzero \\
\addlinespace[2pt]

Trans.-GAN
 & 0.2978 & \dpos{81.36\%}{81} & \dpos{49.53\%}{49}
 & 0.0493 & \dpos{31.29\%}{31} & \dpos{37.58\%}{37}
 & 0.0300 & \dpos{25.36\%}{25} & \dneg{-6.74\%}{6}
 & 0.0255 & \dpos{37.27\%}{37} & \dpos{8.13\%}{8} \\

\quad + $C$
 & 0.1503 & \dpos{63.06\%}{63} & \dzero
 & 0.0308 & \dneg{-10.07\%}{10}& \dzero
 & 0.0320 & \dpos{30.08\%}{30} & \dzero
 & 0.0234 & \dpos{31.72\%}{31} & \dzero \\
\addlinespace[2pt]

DiT
 & 0.1268 & \dpos{56.22\%}{56} & \dpos{37.95\%}{37}
 & 0.0198 & \dneg{-70.86\%}{70} & \dneg{-241.63\%}{241}
 & 0.0291 & \dpos{23.30\%}{23} & \dneg{-13.28\%}{13}
 & 0.0233 & \dpos{31.37\%}{31} & \dneg{-13.72\%}{13} \\

\quad + $C$
 & 0.0787 & \dpos{29.44\%}{29} & \dzero
 & 0.0677 & \dpos{49.99\%}{49} & \dzero
 & 0.0330 & \dpos{32.29\%}{32} & \dzero
 & 0.0265 & \dpos{39.65\%}{39} & \dzero \\
\addlinespace[2pt]

CSDI
 & 0.1119 & \dpos{50.36\%}{50} & \dpos{2.15\%}{2}
 & 0.2198 & \dpos{84.60\%}{84} & \dneg{-1.85\%}{1}
 & 0.0755 & \dpos{70.39\%}{70} & \dneg{-2.12\%}{2}
 & 0.0531 & \dpos{69.90\%}{69} & \dneg{-3.06\%}{3} \\

\quad + $C$
 & 0.1094 & \dpos{49.27\%}{49} & \dzero
 & 0.2238 & \dpos{84.88\%}{84} & \dzero
 & 0.0771 & \dpos{71.01\%}{71} & \dzero
 & 0.0548 & \dpos{70.79\%}{70} & \dzero \\
\addlinespace[2pt]

keGAN
 & 0.6895 & \dpos{91.95\%}{91} & \dpos{41.18\%}{41}
 & 0.0660 & \dpos{48.69\%}{48} & \dneg{-76.21\%}{76}
 & 0.3281 & \dpos{93.19\%}{93} & \dpos{91.88\%}{91}
 & 0.3217 & \dpos{95.03\%}{95} & \dpos{93.15\%}{93} \\

\quad + $C$
 & 0.4056 & \dpos{86.31\%}{86} & \dzero
 & 0.1162 & \dpos{70.88\%}{70} & \dzero
 & 0.0266 & \dpos{16.08\%}{16} & \dzero
 & 0.0220 & \dpos{27.36\%}{27} & \dzero \\

\midrule
\rowcolor{blue!10}
\textbf{Ours}
 & \textbf{0.0555} & \dzero & \dzero
 & \textbf{0.0338} & \dzero & \dzero
 & \textbf{0.0224} & \dzero & \dzero
 & \textbf{0.0160} & \dzero & \dzero \\
\bottomrule
\end{tabular}
\end{table*}

\subsection{Traffic Generation Performance}
Table~\ref{main_table} shows the experimental results of the proposed model and the baseline model on the task of cellular traffic generation. 
Models without '+ $C$' take only noise as input, while '+ $C$' indicates that the model is additionally conditioned on the multimodal spatial representation.
All ‘+$C$' conditional models shared the same train epochs (i.e., 100) for fair comparison.
We can observe that NetSpatial achieves the lowest values for JSD, RMSE, and MAE, indicating that the generated traffic sequences are closer to the real data distribution and more accurate in absolute traffic magnitude. 
Although some diffusion-based models, such as DiT and CSDI, achieve slightly better results on JSD\_DIFF, this metric reflects short-term fluctuations between adjacent time steps rather than long-term temporal structure. 
NetSpatial instead models daily and weekly periodic patterns explicitly and separates them from stochastic residual dynamics through a hierarchical generation process, which better reflects the temporal characteristics of cellular traffic. 

Another observation from Table~\ref{main_table} is that the effect of spatial conditioning varies across different baseline architectures. 
The multimodal spatial representation contains useful information about the relationship between urban context and traffic demand, but its utility depends on the generation framework. For baselines not originally designed to incorporate hierarchical semantic conditions, simply concatenating the spatial embedding with the input noise does not necessarily yield consistent improvements. Under the fixed training budget of 100 epochs used in our experiments, the added condition may improve distribution-level similarity while not always reducing pointwise errors such as RMSE or MAE, a pattern that can be observed in models such as RNN-GAN and several other baselines. In contrast, models whose architectures already integrate urban semantic information tend to benefit more from additional spatial conditions. For example, keGAN incorporates an urban knowledge graph to model hierarchical urban features, making its design more compatible with external spatial representations and resulting in larger improvements in RMSE and MAE when conditioning information is introduced.

\begin{figure}[t]
  \centering
  \includegraphics[width=\linewidth]{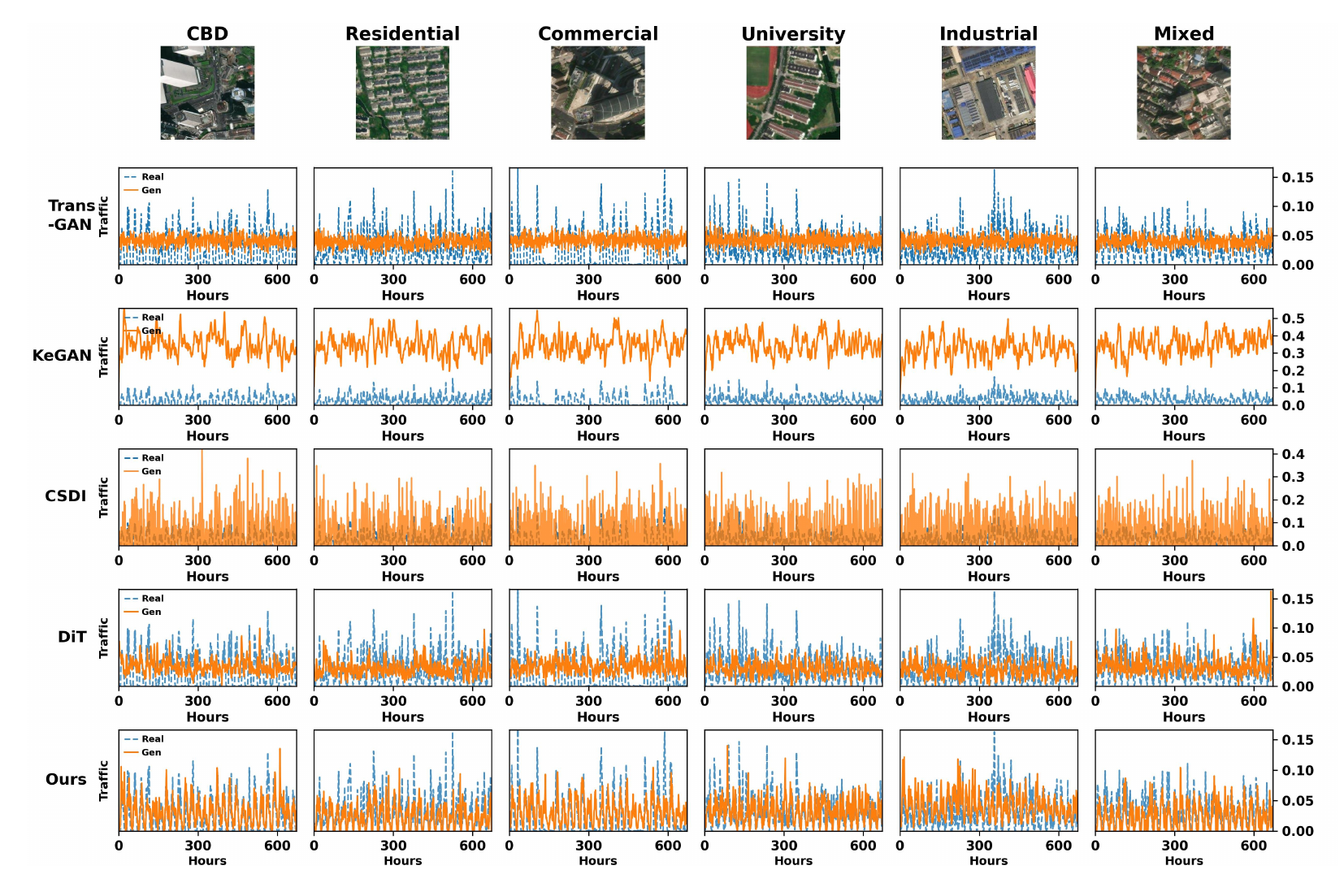}
  \caption{Visualization of generated BS traffic at a monthly granularity.}
  \label{fig_month}
\end{figure}

\begin{figure}[t]
  \centering
  \includegraphics[width=\linewidth]{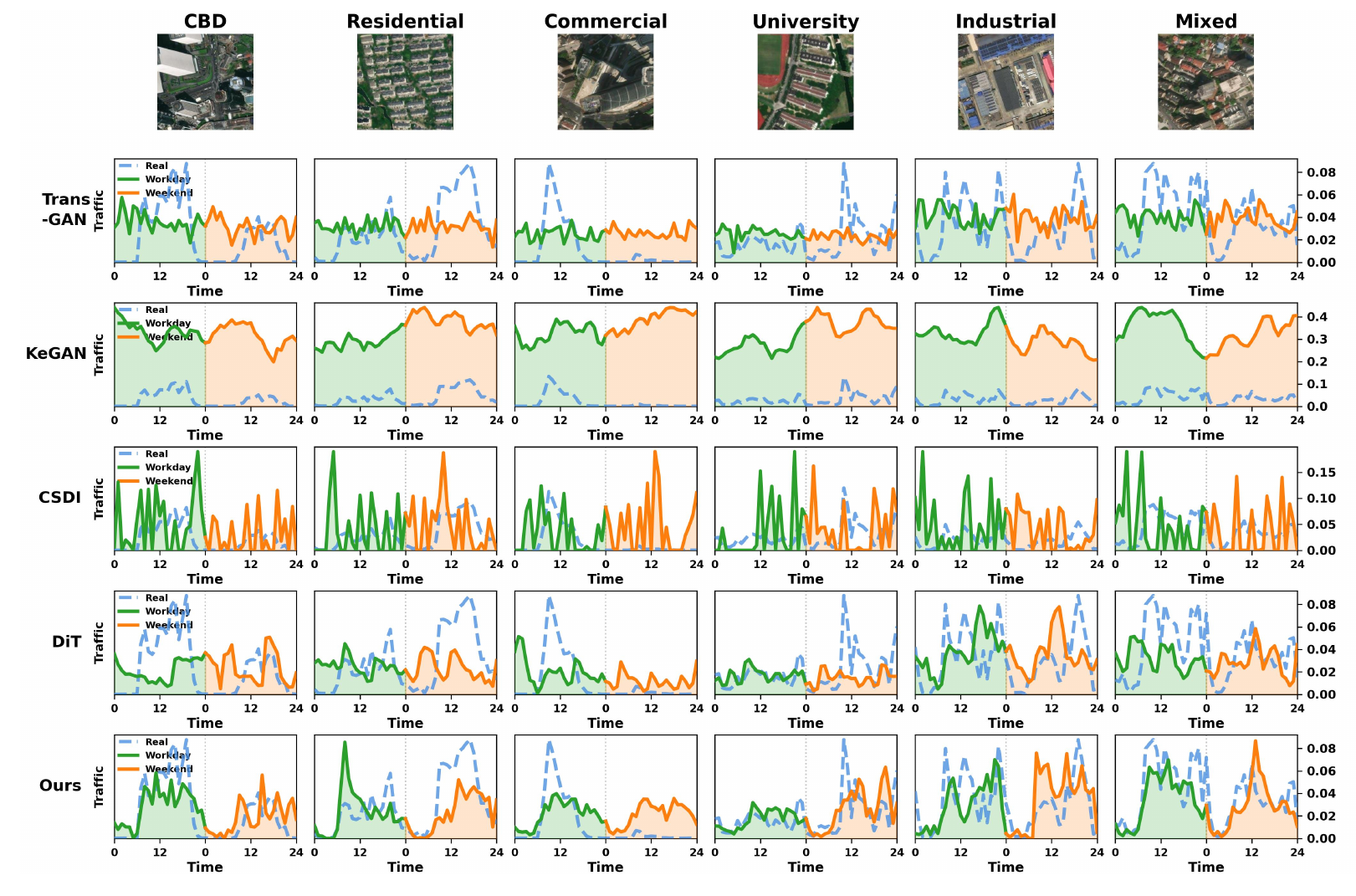}
  \caption{Visualization of generated BS traffic at a daily granularity.}
  \label{fig2_day}
\end{figure}

We also conduct a case study within the corresponding spatial functional zones, selecting six functional regions, i.e., central business district (CBD), residential area, commercial area, university area, industrial area, and mixed area, and compare the generated traffic and the real one across different functional regions. 
As shown in Figures~\ref{fig_month} and~\ref{fig2_day}, we observe clear daily and monthly temporal patterns in the generated traffic data.
This verifies that the multimodal spatial semantics are successfully synthesized into high-fidelity traffic patterns across heterogeneous urban zones, demonstrating the effectiveness of the multimodal perception and fusion architecture and the multimodal conditional generation mechanism.

\subsection{Visualization Platform for Cellular Planning and Operations}
We develop a visualization platform for NetSpatial, built on Mapbox, with selected functions integrated into a web-based application for demonstration.  
The demonstration video is available at \url{https://anonymous.4open.science/r/Traffic-Flow-Predictor-Demo}.
The system features BS analytics, enabling traffic visualization and site selection.

\subsubsection{What-if Planning Under Spatial Dynamics.}

\begin{figure}[t]
  \centering
  \includegraphics[width=\linewidth]{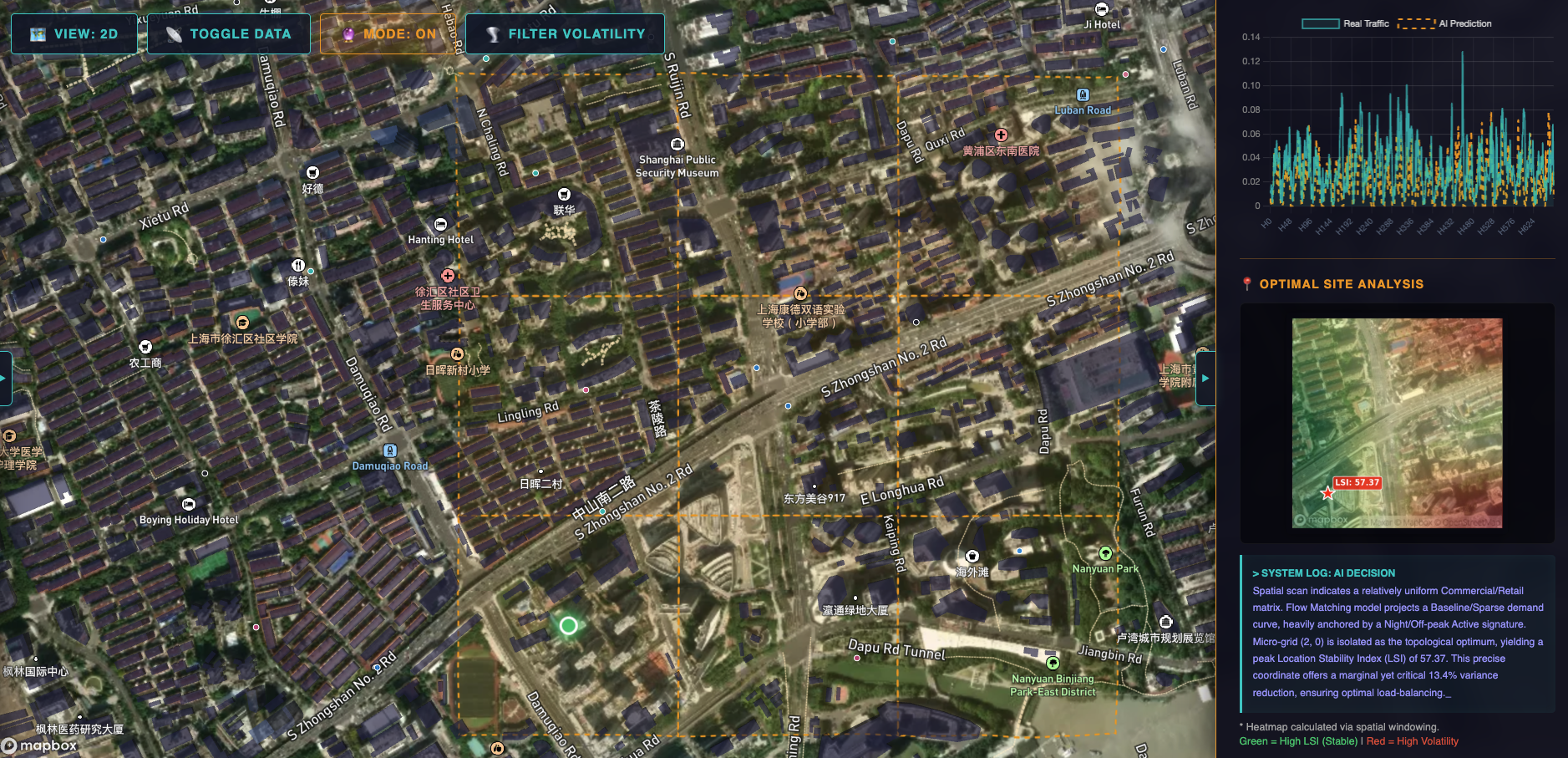}
  \caption{User interface of our BS Analytics Module.}
  \label{fig_what_if}
\end{figure}

Our platform presents the BS siting capability from a spatial context perspective, offering an interactive experience. As illustrated in Figure~\ref{fig_what_if}, users can explore a 3 km × 3 km urban space where our conditional generative model estimates the traffic sequence associated with each feasible grid if a BS were deployed there. The platform is designed to support different deployment objectives by scoring and ranking candidate grids according to the selected utility. In the current demonstration, we use the LSI as the example, which highlights locations with relatively high and temporally stable traffic demand. To make the decision process more transparent, the system further leverages a large language model to generate situational explanatory rationales. Beyond siting, the platform also supports switching between 2D and 3D views for multi-angle visualization of BS traffic distribution.
% In summary, this extended module of NetSpatial provides users with a novel way to evaluate urban connectivity, combining spatial intelligence with interpretable reasoning to enhance accessibility to complex network deployment data.

\subsubsection{What-to-Do Operations Under Volatility.}
\begin{figure*}[t]
  \centering
  \includegraphics[width=\linewidth]{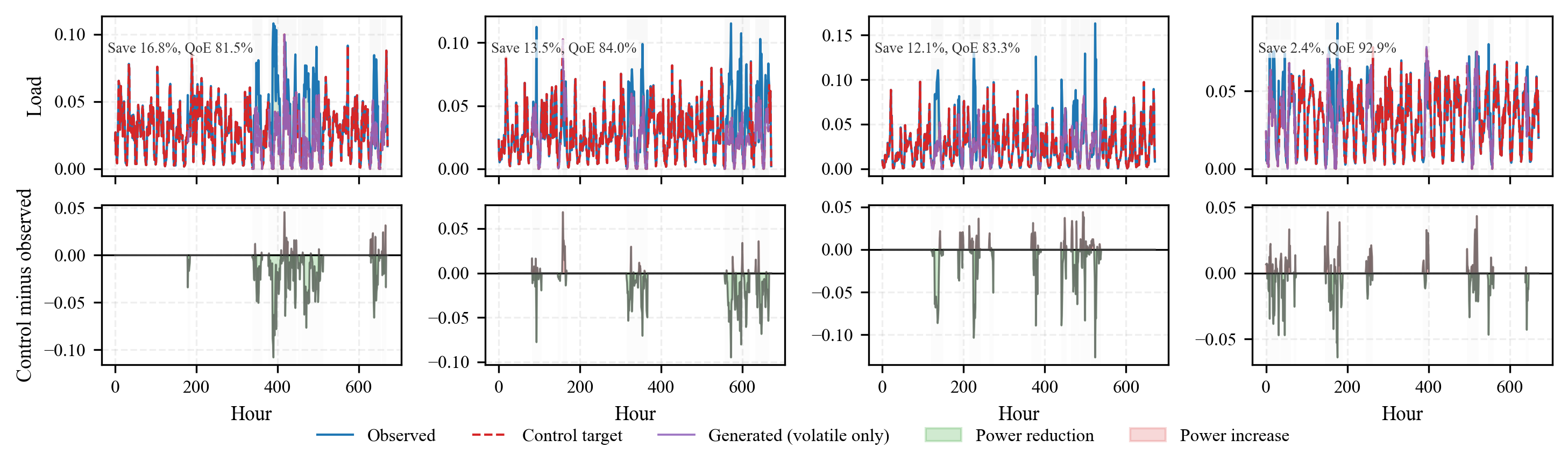}
  \caption{Decision-centric control visualization for four BSs. Top row: observed load, control target, and generated signals used only during high-volatility windows. Bottom row: power adjustment signal showing regions of energy reduction and increase.}
  \label{fig:what_to_do_control}
\end{figure*}

We evaluate the system's ability to make what-to-do decisions for energy control at the BS. At each station, the system identifies high-volatility intervals and activates only the generated targets within those windows, while preserving observed traffic elsewhere. As shown in Fig.~\ref{fig:what_to_do_control}, the top row visualizes observed load, the control target, and the generated signal used during volatile periods. The bottom row shows the corresponding power adjustment signal, highlighting energy reduction and increase segments. 
% We define QoE as the demand satisfaction ratio (1 - unmet\_demand/total\_demand), measure energy savings against a no-control baseline, and assume transmit power scales linearly with traffic load.

Let $\rho_t^{\mathrm{obs}}$ and $\rho_t^{\mathrm{ctrl}}$ denote the normalized
traffic load at time slot~$t$ without and with control, respectively.
Following the EARTH linear power model~\cite{auer2011much}, the idle power
cancels in relative comparisons, yielding the energy saving ratio
\begin{equation}\label{eq:energy-saving}
  \eta = 1 - {\sum\nolimits_{t=1}^{T} \rho_t^{\mathrm{ctrl}}}
             \big/{\sum\nolimits_{t=1}^{T} \rho_t^{\mathrm{obs}}}\,.
\end{equation}
Service quality is measured by the demand-weighted satisfaction
ratio~\cite{oh2013dynamic}, which only penalises under-provisioning,
\begin{equation}\label{eq:qoe}
  \mathrm{QoE} = 1 - {\sum\nolimits_{t=1}^{T}
    [\rho_t^{\mathrm{obs}} - \rho_t^{\mathrm{ctrl}}]^{+}}
    \big/{\sum\nolimits_{t=1}^{T} \rho_t^{\mathrm{obs}}}\,,
\end{equation}
where $[\cdot]^{+}=\max(\cdot,0)$.

The results indicate that the platform makes consistent and interpretable what to do decisions, activating generated control only when volatility is high and maintaining stable behavior elsewhere. Under this selective activation strategy, the model achieves up to 16.8\% energy savings while retaining over 80\% QoE, demonstrating an effective trade-off between energy efficiency and service quality.

\begin{figure*}[t]
  \centering
  \includegraphics[width=\linewidth]{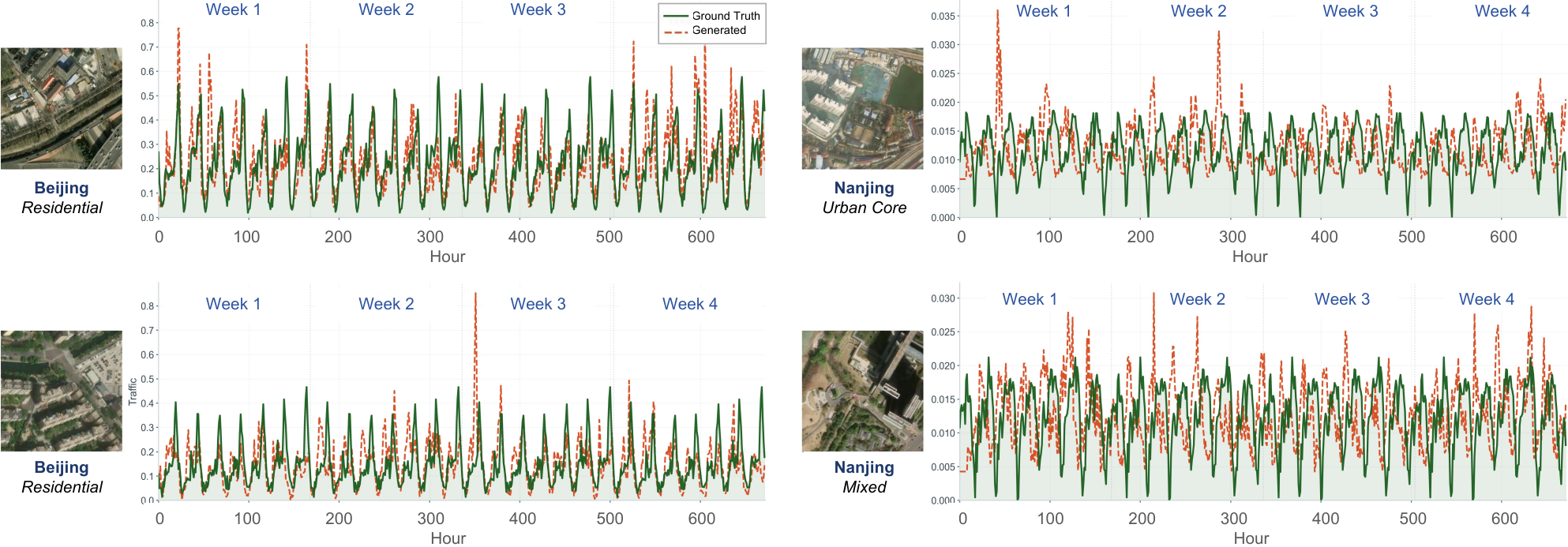}
  \caption{Cross City Zero Shot Transfer Results.}
  \label{fig_cross}
\end{figure*}

\subsection{Cross-City Robustness Testing}
To evaluate robustness and generalization, we conduct cross-city zero-shot transfer experiments where the model trained on Shanghai is directly applied to Beijing and Nanjing without fine-tuning. Evaluation is performed on 50 BSs in Beijing covering diverse urban regions, and 50 BSs in Nanjing spanning Urban Core and Mixed districts.
Figure~\ref{fig_cross} shows four randomly selected BSs as examples for visualization. 
For Beijing BSs, the model achieves a best correlation of 0.5738 and a mean MAE of 0.1071, successfully capturing weekly periodicity, daily fluctuations, and peak characteristics aligned with ground truth. 
For Nanjing BSs, the model maintains consistent zero-shot generation with a mean MAE of 0.0210 across 50 BSs, demonstrating accurate traffic volume estimation and robustness across different urban characteristics. 
Importantly, following the evaluation framework from prior cross-city traffic prediction studies~\cite{zhang2023deep}, our zero-shot approach enables direct deployment in new cities using only satellite imagery and geographic information without requiring historical traffic data, showing that NetSpatial learns generalizable spatiotemporal representations rather than city-specific patterns.

\subsection{Ablation Study}
We next conduct ablation studies to evaluate the effectiveness of the position encoder within the multi-level conditional generation model and of multimodal representation learning for cellular traffic generation.

\textbf{Effect of the position encoder}. 
Our model employs a three-layer conditional generation structure to explicitly model time series with different characteristics, including periodic and non‑periodic components, and leverages multimodal spatial representations to guide cellular traffic generation. To test the effect of introducing a position encoder to capture non‑periodic fluctuations, meaning the timing of the fluctuations and their corresponding amplitudes, from multimodal spatial representations, we conduct an ablation study by removing the position encoder, allowing the lowest-layer module to generate cellular traffic without this non‑periodic information. The experimental results are shown in Table~\ref{tab:ablation-delta}. We observe that removing the position encoder reliably degrades our model's performance across all metrics, with some metrics dropping by more than 50\%.

\textbf{Effect of the multimodal representation learning}. 
We adopt a generative mask-modeling strategy for multimodal representation learning. To assess its contribution to spatial embeddings, we perform two ablation studies, i.e., removing the conditioning and replacing it with embeddings obtained via a CLIP-based approach. The experimental results are shown in Table~\ref{tab:ablation-delta}.
We observe that both ablation variants degrade our model's performance across most metrics, with some metrics dropping by over 65\%.

\begin{table}[t]
\centering
\caption{Ablation study. $\Delta$: relative improvement of Ours over each variant. Lower is better. Cell color: \colorbox{pos3}{\textcolor{white}{green}} = Ours better; \colorbox{neg3}{\textcolor{white}{red}} = Ours worse.}
\label{tab:ablation-delta}
\setlength{\tabcolsep}{0.8pt}
\renewcommand{\arraystretch}{1.1}
\scriptsize
\begin{tabular}{l cc cc cc cc cc}
\toprule
\rowcolor{gray!15}
\textbf{Method} &
\multicolumn{2}{c}{\textbf{JSD}} &
\multicolumn{2}{c}{\textbf{JSD\_D}} &
\multicolumn{2}{c}{\textbf{RMSE}} &
\multicolumn{2}{c}{\textbf{MAE}} &
\multicolumn{2}{c}{\textbf{T/ms}} \\
\cmidrule(lr){2-3}\cmidrule(lr){4-5}\cmidrule(lr){6-7}\cmidrule(lr){8-9}\cmidrule(lr){10-11}
\rowcolor{gray!15}
 & Val & $\Delta$
 & Val & $\Delta$
 & Val & $\Delta$
 & Val & $\Delta$
 & Val & $\Delta$ \\
\midrule
w/o pos.
 & 0.1801 & \dpos{69.17\%}{69}
 & 0.0718 & \dpos{52.86\%}{52}
 & 0.1459 & \dpos{84.67\%}{84}
 & 0.1026 & \dpos{84.40\%}{84}
 & 86.70 & \dneg{-1.57\%}{1} \\

w/o cond.
 & 0.0799 & \dpos{30.53\%}{30}
 & 0.0058 & \dneg{-486.54\%}{500}
 & 0.0516 & \dpos{56.72\%}{56}
 & 0.0404 & \dpos{60.40\%}{60}
 & 89.80 & \dpos{1.94\%}{1} \\

clip cond.
 & 0.1626 & \dpos{65.86\%}{65}
 & 0.0239 & \dneg{-41.84\%}{41}
 & 0.0274 & \dpos{18.33\%}{18}
 & 0.0194 & \dpos{17.46\%}{17}
 & 93.60 & \dpos{5.92\%}{5} \\

\midrule
\rowcolor{blue!10}
\textbf{Ours}
 & \textbf{0.0555} & \dzero
 & \textbf{0.0338} & \dzero
 & \textbf{0.0224} & \dzero
 & \textbf{0.0160} & \dzero
 & \textbf{88.06} & \dzero \\
\bottomrule
\end{tabular}
\end{table}

\begin{figure}[t]
  \centering
  \includegraphics[width=\linewidth]{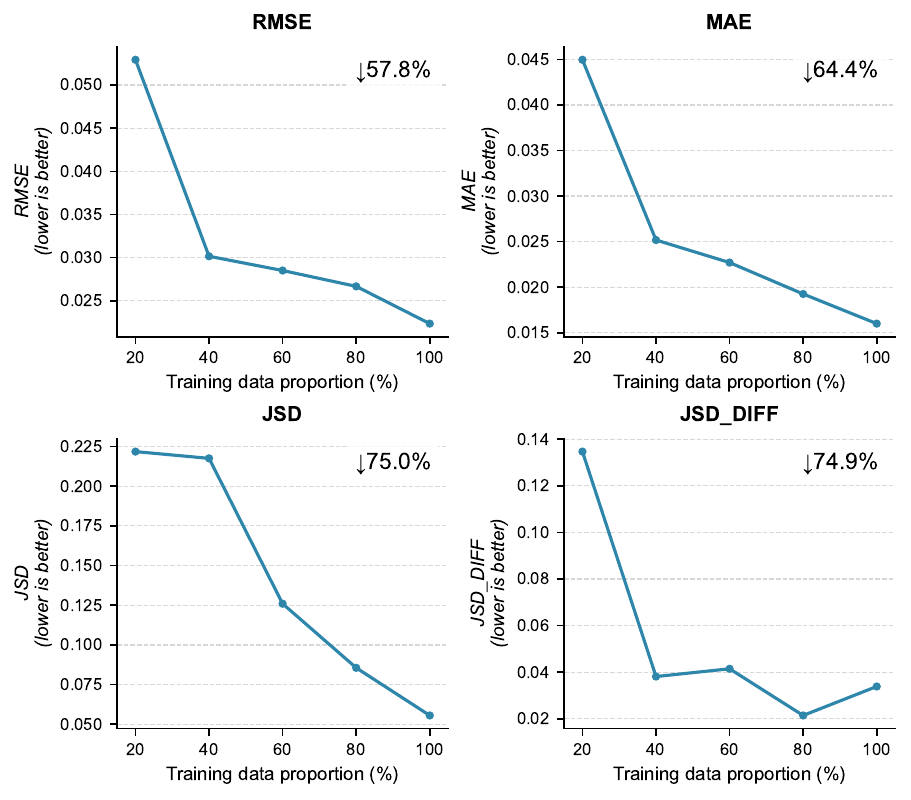}
  \caption{Performance Sensitivity to Training Data Size.}
  \label{fig_ds}
\end{figure}

\textbf{Sensitivity of training data size}.
We test how \textsc{NetSpatial}'s performance changes with training data size by varying the proportion of BSs. 
As shown in Figure~\ref{fig_ds}, performance improves consistently as more data becomes available. 
Notably, gains are most dramatic when increasing from 20\% to 40\%, followed by a plateau, and then a renewed improvement beyond 80\%. 
These findings indicate that our unsupervised temporal clustering generalizes well without overfitting. 
Practically, if collecting $>$80\% of data is infeasible, targeting $\sim$40\% offers the best cost-effectiveness.

\section{Related Work}
\textbf{Network Traffic Generation.}
Early traffic generation relied on mathematical models such as Poisson processes and Markov chains~\cite{weigle2006tmix, vishwanath2009swing, zhang2015survey}, while later autoregressive models~\cite{xu2020stan, cardoso2019generation} were introduced to capture temporal dependencies. GANs have been widely adopted for this task. PcapGAN~\cite{dowoo2019pcapgan} generates packet-level traces, while DoppelGANger~\cite{lin2019generating, yin2022practical} jointly models attributes and time series to improve fidelity. Diffusion probabilistic models have recently emerged as an alternative generative paradigm~\cite{yang2024survey}. NetDiffusion~\cite{jiang2024netdiffusion} and NetDiffus~\cite{sivaroopan2024netdiffus} apply diffusion models to protocol-constrained packet trace synthesis. For spatio-temporal traffic modeling, DiffSTG~\cite{wen2023diffstg} extends denoising diffusion to spatio-temporal graphs, and DiffUFlow~\cite{zheng2023diffuflow} combines spatio-temporal and semantic features for urban flow inference. However, these approaches have notable limitations for city-level cellular traffic generation. Packet-level methods~\cite{ring2019flow, dowoo2019pcapgan, jiang2024netdiffusion, zhang2025packetdiff} do not scale beyond individual BSs. Forecasting and super-resolution methods~\cite{wen2023diffstg, zheng2023diffuflow} require existing observations and cannot generate traffic from scratch. Moreover, none of these methods incorporates multimodal spatial conditioning, limiting their applicability to spatially aware synthesis for network planning and what-if analysis.

\textbf{Cellular Planning and Operations.}
Traditional cellular network planning relies on empirical propagation  models~\cite{rappaport2010wireless, rappaport2017overview} and  optimization-based approaches for BS placement~\cite{kalantari2024optimal, su2025gennet, su2025jointly, hassan2024spaceris}. 
Recent works apply machine learning to predict radio maps from building geometry and aerial images~\cite{jiang2025unirm, feng2025recent}.
Digital twin frameworks~\cite{rodrigo2023digital, ma2023adaptive, liu2024digital} integrate simulation with real-world data for network planning and what-if analysis. 
However, existing tools cannot synthesize realistic city-scale traffic distributions conditioned on spatial context, limiting their utility for greenfield deployments where ground-truth observations are unavailable.
\section{Discussion}
NetSpatial provides a unified system that links network planning and operation through spatially conditioned traffic generation. By leveraging multimodal spatial context, the system enables scenario-based analysis of cellular networks and can estimate traffic demand under different deployment and operational conditions. This generative capability supports both deployment planning, by evaluating candidate locations using spatial context, and network operation, by producing traffic forecasts that guide strategies such as sleep scheduling and load balancing.

Despite these advantages, several limitations remain. First, the current system relies on static spatial features such as satellite imagery and POI distributions, while dynamic factors including mobility patterns and special events may also influence traffic demand. Second, the model generates traffic conditioned on spatial context but does not explicitly model interactions among neighboring base stations, which may affect load redistribution in dense deployments. Future work could incorporate mobility data and network topology information to further improve demand modeling and operational decision support.

\section{Conclusion}
We presented \textbf{NetSpatial}, a spatially conditional system for cellular traffic generation that supports both cellular planning and operation. By leveraging multimodal spatial context, NetSpatial enables the generation of realistic traffic patterns under diverse spatial scenarios, allowing operators to evaluate candidate deployments and support operational decisions such as traffic forecasting and load management. Experiments on real-world cellular datasets demonstrate that the proposed system effectively captures spatial-temporal traffic characteristics and provides a practical tool for data-driven planning and optimization.
Future work will focus on extending NetSpatial along three directions. First, incorporating richer spatial signals and urban structural information could improve spatial semantics and traffic representation. Second, evaluating the model across broader multi-city datasets would help study generalization and regional variability. Third, integrating additional network measurements and developing online modeling capabilities could further support dynamic and service-aware network management.

\newpage
\bibliographystyle{ACM-Reference-Format}
\bibliography{Reference}

\end{document}